\magnification=1100

\hsize 17truecm
\vsize 23truecm

\font\twelvec=msbm10 at 12pt
\font\sevenc=msbm10 at 9pt
\font\fivec=msbm10 at 7pt

\newfam\co
\textfont\co=\twelvec
\scriptfont\co=\sevenc
\scriptscriptfont\co=\fivec

\def\Const{\mathop{\rm Const.}\nolimits}
\def\det{\mathop{\rm det}\nolimits}
\def\exp{\mathop{\rm exp}\nolimits}

\def\ext{\mathop{\rm ext}\nolimits}

\def\im{\mathop{\rm Im}\nolimits}

\def\Gram{\mathop{\rm Gram}\nolimits}

\def\ker{\mathop{\rm Ker}\nolimits}
\def\lim{\mathop{\rm lim}\nolimits}

\def\Ran{\mathop{\rm Ran}\nolimits}

\def\supp{\mathop{\rm supp}\nolimits}

\def\inf{\mathop{\rm inf}\nolimits}
\def\sgn{\mathop{\rm sgn}\nolimits}

\def\neigh{\mathop{\rm neigh}\nolimits}

\def\Sum{\displaystyle\sum}
\def\e{\mathop{\rm \varepsilon}\nolimits}

\baselineskip 15pt

\centerline {\bf ANDREEV REFLECTION}

\centerline{\bf AND THE SEMICLASSICAL BOGOLIUBOV-DE GENNES HAMILTONIAN:}

\centerline{\bf RESONANT STATES}
\bigskip
\centerline{A. BENSOUISSI ${}^{1,2}$, N. M'HADBI ${}^{2}$ {\it\&} M. ROULEUX ${}^{1}$}
\bigskip
\centerline {${}^{1}$ Universit\'e du Sud Toulon-Var, and Centre de Physique Th\'eorique}

\centerline {CPT, Case 907, 13288 Marseille Cedex 9, France}

\centerline {${}^{2}$ Universit\'e de Tunis El-Manar, D\'epartement de Math\'ematiques, 1091 Tunis, Tunisia}
\bigskip
\noindent {\bf Abstract}: 
We present a semi-classical analysis of the opening of superchannels in gated mesoscopic SNS junctions.
For perfect junctions (i.e. hard-wall potential), this was considered by [ChLeBl] in the framework of scattering matrices. 
Here we allow for imperfections in the junction, so that the complex order parameter continues as a smooth function, which is a
constant in the superconducting banks, and vanishes rapidly inside the lead. We obtain
quantization rules for resonant Andreev states near energy $E$ close to the Fermi level, including the determination of the resonance
width.
\medskip
\noindent{\bf 0. Introduction}.
\smallskip
Bogoliubov-de Gennes Hamiltonian is a $2\times2$ matrix ${\cal P}(x,\xi)$ defined for $(x,\xi)\in T^*{\bf R}$,
which describes the dynamics of a pair of quasi-particles (hole/electron) in a 1-D metallic lead connecting 
2 superconducting contacts. Diagonal terms are of the form $\pm(\xi^2-\mu(x))$, where $\mu(x)$ stands for the  chemical 
potential, while
the off-diagonal interaction with the supraconducting bulk is modeled 
through a complex potential,
or superconducting gap, $\Delta_0 e^{i\phi_\pm/2}$ at the boundary~; due to the finite range of the junction,
we may consider that the interaction continues to a smooth function $x\mapsto\Delta(x)e^{i\phi(x)/2}$ on a neighborhood of the lead
(say, the interval $[-L,L]$.~) So we assume that
$\Delta(x)$, we will call henceforth the ``gap function'', is a smooth positive function, increasing on $x>0$,
with $\Delta(x)=0, |x|\leq x_1<L$ and $\Delta(x)=\Delta_0, |x|\geq x_2>L$ (ignoring the fact that $\Delta$ 
shows typically isolated zeroes (vortices) in the supraconducting bank). In the same way, we will assume that 
$\phi(x)=\sgn(x)\phi$ takes only 2 values.
The chemical potential $\mu(x)$ will be extended also to a smooth positive function on a neighborhood of $[-L,L]$,
constant and $>\Delta_0$ for $|x|\geq x_2$. 
As is usual for a metal, we assume that $\mu$ and $\Delta$ are even in $x$,
which provides this model with the CPT symmetry. 

We introduce a ``Planck constant'' $h>0$, which stands for the ratio of $L$ to the characteristic de Broglie wave-length,
and take usual $h$-Weyl quantization, 
$${\cal P}(x,hD_x)=\pmatrix{(hD_x)^2-\mu(x)\kern 1pt &  \Delta(x)e^{i\phi(x)/2}\kern 1pt \cr
\Delta(x)e^{-i\phi(x)/2}\kern 1pt & -(hD_x)^2+\mu(x)\kern 1pt} 
\leqno(0.1)$$
An electron $e^-$ moving in the metallic lead with energy $0<E\leq\Delta$ (measured with respect to Fermi level $E_F$)
and kinetic energy $K_+(x)=\mu(x)+\sqrt{E^2-\Delta(x)^2}$ is 
reflected back from the supraconductor as a hole $e^+$, with kinetic energy $K_-(x)=\mu(x)-\sqrt{E^2-\Delta(x)^2}$,
injecting a Cooper pair into the bulk. 
When $\inf _{[-L,L]}\mu(x)\geq E$, and $\phi\neq0$,  
this process yields so called phase-sensitive Andreev states, carrying supercurrents proportional 
to the $\phi$-derivative of the eigen-energies $E_k(h)$ of ${\cal P}(x,hD_x)$.
Since ${\cal P}(x,hD_x)$ is self-adjoint,
there is of course also an electron moving to the left, and a hole
moving to the right (in fact, ${\cal P}(x,hD_x)$ is the Hamiltonian for 2 pairs of quasi-particles), for
no net transfer of charge can occur through the lead in absence of thermalisation. 
So we stress that Bogoliubov-de Gennes Hamiltonian is only a simplified model for superconductivity, 
and that a more thorough treatment should also take into account 
the self-consistency relations coupling the quasi-particle
with the gap function $\Delta(x)$ and the phase $\phi(x)$, that we treat here as ``effective potentials''
(see [KeSo]).

In the case where $\Delta(x)$ 
is a ``hard-wall'' potential, this was studied in [ChLesBl], [CaMo]
in the framework of scattering matrices.
In [BeIfaRo] , we derived semi-classical quantization rules for Andreev states near energy $E$,
from a microlocal study of the Hamiltonian in the ``inner region'' $\Delta(x)\leq E$ alone.
For simplicity, we assumed that $\Delta(x)$ varies linearly near $E$, namely if 
$x_0\in]x_1,x_2[$ is such that $\Delta(x_0)=E$, then $\mu(x)=\mu=\Const $  and 
$\Delta(x)=E+\alpha(x-x_0)$ near $x_0$. 

Here we want to take also into account the ``outer region'' $\Delta(x)\geq E$ (i.e. $|x|\geq x_0$) of the junction, 
entering the supraconducting bulk.
As a matter of fact, the microlocal solutions, purely oscillating in $\Delta(x)\leq E$, acquire a complex phase in $\Delta(x)\geq E$,
which is of course related to phase-space tunneling. 
We make the assumption that the junction is extended, in such a way that the quasi-particle turns into a resonant state before 
creating a new Cooper pair, its dynamics still being governed by Bogoliubov-de Gennes Hamiltonian. 
So we assume that $\mu(x)$ and $\Delta(x)$ are defined on the entire real line, taking constant values for $|x|\geq x_2>L$, so that 
(0.1) can be defined as a self-adjoint operator on $L^2({\bf R})\otimes{\bf C}^2$. We will translate the usual theory of analytic dilations [ReSi]
in the context of CPT symmetry, and find semi-classical resonances
near a ``scattering'' Andreev level,
i.e. complex correction to the real eigen-energies $E_k(h)$ of ${\cal P}(x,hD_x)$. 
\medskip
\noindent{\bf 1) The real part of the resonances}.
\smallskip
The bicharacteristic set in $\{\xi>0\}$ at energy $E$, of the form $\det{\cal P}(x,\xi)-E=0$, or
$\xi^2=K_\pm(x)$, consists of~: (1) two real curves $\rho_\pm$ 
over $[-x_0,x_0]$, joining smoothly to a close curve at the ``branching points'' $a'=(-x_0,\xi_0)$ and $a=(x_0,\xi_0)$
(so to make $\rho_+\cup\rho_-$ diffeomorphic
to ${\bf S}^1$)~;
(2) complex branches $\rho^<_\pm$ over $]-\infty,-x_0]$, and $\rho^>_\pm$ over $[x_0,+\infty[$ respectively. 
They all have a vertical tangent
at $a,a'$. We complete this picture by reflection on the $x$ axis, denoting the 
corresponding branching points by $b',b$.
\medskip
\noindent{\bf a) Microlocal solutions supported on $\rho_\pm$}
\smallskip
First we recall from [3] the construction of distributions
microlocalized on the Lagrangians $\rho_\pm$, and verifying the PT symmetries of the problem. 
We denote the parity operator by ${}^\vee:u(x)\to u(-x)$, and the time reversal operator by
${\cal I}:u(x)\to\overline{u(x)}$.
\medskip
\noindent {\bf Definition 1.1}: {\it We call ``admissible ${\bf C}^2$-valued Lagrangian distribution'' an oscillatory integral
$$I(S,\varphi)(x,h)=(2\pi h)^{-d/2}\int_{{\bf R}^d} e^{i\varphi(x,\Theta,h)/h}S(x,\Theta;h)d\Theta\leqno(1.1)$$
with the following properties~: (1) $\varphi(x,\Theta,h)$ denotes
a non degenerate phase-function, and 
$$S(x,\Theta;h)=S_0(x,\Theta;h)+hS_1(x,\Theta;h)+\cdots$$
a ${\bf C}^2$-valued  amplitude (i.e. a classical symbol in $h$), $S_0={e^{i\phi/2}X\choose Y}$ possibly depending on $h$ 
(with the property that $\phi(x)=\sgn(x)\phi$)~; (2) The symbols $X=X(x,\Theta,h), Y=Y(x,\Theta,h)$
have their principal part ${X_0\choose Y_0}=\lambda(x,\Theta;h){X'_0\choose Y'_0}$, $\lambda\in{\bf C}$,
proportional to a real vector ${X'_0\choose Y'_0}$, depending also on $(x,\Theta;h)$. }
\smallskip
Of course, all these functions may depend on additional parameters. 
One of the main problem consists in finding microlocal solutions near the branching points $a,a'$.  
Due to PT symmetry, it suffices to focus on
$a=(x_0,\xi_0)$. In $h$-Fourier representation, the Hamiltonian takes the form
$${\cal P}^a(-hD_\xi, \xi)=\pmatrix{\xi^2-\mu\kern 1pt &  e^{i\phi/2}(E-\alpha hD_\xi-\alpha x_0)\kern 1pt \cr
e^{-i\phi/2}(E-\alpha hD_\xi-\alpha x_0)\kern 1pt & -\xi^2+\mu\kern 1pt } 
$$
where $\mu=\xi_0^2$ is a constant, equal to the value of the chemical potential at $x_0$. Consider the equation 
$({\cal P}^a(-hD_\xi,\xi)-E )\widehat U=0$, where $\widehat U=\pmatrix{\widehat\varphi_1\cr \widehat\varphi_2}$. 
Clearly, the system decouples, and
to account for time-reversal symmetry, it is convenient
to introduce the scaling parameter $\beta=\sqrt\alpha(2\xi_0)^{-3/2}$, 
together with the changes of variables $\xi=\xi_0(\pm2\beta\xi'+1)$.
The functions $\widetilde u_{\pm\beta}(\xi')=(\xi^2-\mu-E )^{-1/2}e^{-i(E-\alpha x_0)\xi/\alpha h}\widehat\varphi_2$
satisfy a second order ODE of the form
$$(\widetilde P_{\pm\beta}(-hD_{\xi'},\xi',h)-{E_1^2\over\beta^2})\widetilde u_{\pm\beta}(\xi')=0\leqno(1.2)$$
with $E_1=(2\xi_0)^{-2}E$, and 
$$\eqalign{
\widetilde P_{\pm\beta}&(-hD_{\xi'},\xi',h)=(hD_{\xi'})^2+ \bigl(\xi'\pm\beta\xi'^2\bigr)^2\cr
&+h^2(2\xi_0)^{-2}\beta^2
\bigl(2\beta^2\xi'^2\pm2\beta\xi'+{3\over4}+E_1 \bigr)
\bigl(\beta^2\xi'^2+\beta\xi'-E_1 )^{-2}\cr
}\leqno(1.3)$$
Operators $\widetilde P_{\beta}$ and $\widetilde P_{-\beta}$ are unitarily equivalent, and so have the same spectrum. 
Up to the ${\cal O}(h^2)$ term, $\widetilde P_{\pm\beta}(-hD_{\xi'},\xi',h)$ have the structure of an ``anharmonic oscillator",
with ``potential wells'' at $\xi'=0,\mp1/\beta$ separated by a ``barrier'' at $\xi'=\mp1/(2\beta)$. 
It is also well known [HeSj] that, viewed as a  $h$-PDO of order 0, microlocally defined near $(x',\xi')=0$,
$\widetilde P=\widetilde P_{\pm\beta}$ can be taken to the normal form
of a harmonic oscillator, away from the ``barrier''.
More precisely, there exists a real-valued analytic symbol 
$F(t,h)=F_{\pm\beta}(t,h)\sim\Sum_{j=0}^\infty F_j(t)h^j$,
defined for $t\in\neigh (0)$, $F_0(0)=0$, $F'_0(0)={1\over2}$, $F_1(t)=\Const $, and 
(formally) unitary FIO's $A=A_{\pm\beta}$ 
whose canonical transformations $\kappa_A$ 
defined in a neighborhood of (0,0), are close to identity 
and map this point onto itself, such that 
$$A^*F(\widetilde P,h)A=P_0={1\over 2}((hD_\eta)^2+\eta^2-h)$$
Define the large parameter $\nu$ by $F({E^2\over2\xi_0\alpha},h)=\nu h$.
So $\widetilde u=\widetilde u_{\pm}$ solves (1.2) microlocally near (0,0) iff $v=A^*\widetilde u$ solves Weber equation 
$(P_0-\nu h)v=0$
microlocally near (0,0), when $\nu h\sim{E^2\over4\xi_0\alpha}$ is small enough.
The well known parabolic cylinder functions $D_\nu$ and $D_{-\nu-1}$,
provide with a basis of solutions of ${1\over 2}((hD_\eta)^2+\eta^2-h)v=\nu v$. We shall use $D_{-\nu-1}$, and write 
$$v=A^*\widetilde u_{\pm\beta}=
\Sum_{\e =\pm1}\alpha^{(-\nu-1)}_{\e ,\pm\beta} D_{-\nu-1}(i\e (h/2)^{-1/2}\eta)\leqno(1.4)$$
for complex constants $\alpha^{(-\nu-1)}_{\e ,\pm\beta}$. 

These microlocal solutions can be expressed in the spatial representation by taking inverse $h$-Fourier transformation~;
expanding integrals of the type (1.4) by stationary phase, both pieces of bicharacteristics $\rho_\pm$ contribute to $U_{\e ,\pm\beta}$
near $a$. Microlocal solutions near $a'$ are deduced by PT symmetry. 

Once microlocal solutions $U^{a,-\nu-1}_{\e ,\beta}$ have been obtained that way near the branching point $a$, 
it is standard to extend them up to $a'$
as WKB solutions $(U^{a,-\nu-1}_{\e ,\beta})_{\ext }$, taking advantage that ${\cal P}$ has simple characteristics away from $a,a'$. 
When $\Delta(x)\equiv0$, i.e. for 
$-x_1\leq x\leq x_1$, they are completely decoupled, which means that the solution is either a 
pure {\it electronic state}, i.e. colinear to the vector ${1\choose0}$ of ${\bf C}^2$, or 
pure {\it hole state}, i.e. colinear to ${0\choose1}$. Otherwise, they are a superposition of electronic/hole states.
We summarize these constructions in the~:
\medskip
\noindent{\bf Proposition 1.2}: {\it For $x<x_0$ near $x_0$,
there are 2 basis of oscillating microlocal solutions of $({\cal P}^a-E)U=0$ indexed by $\e =\pm1$~:
$$\Sum_{\rho=\pm1}U^{a,\nu}_{\rho,\e ,\pm\beta}(x,h'),\quad \Sum_{\rho=\pm1}U^{a,-\nu-1}_{\rho,\e ,\pm\beta}(x,h')$$
Here the branch with $\rho=\rho_\pm=\pm1$ is microlocalized on $\rho_\pm$, i.e.
the part on $\rho_+$ ($\xi>\xi_0$ near $a$),
belongs to the electron state, while the part $\rho_-$ ($\xi<\xi_0$ near $a$) belongs to the hole state~; they 
satisfy, for $\rho=\pm$~:
$$U^{a,-\nu-1}_{\rho,-,\beta}=U^{a,-\nu-1}_{\rho,+,\beta}+{\cal O}(h)\leqno(1.5)$$
and
$$U^{a,-\nu-1}_{-,\e ,\beta}=U^{a,-\nu-1}_{+,\e ,-\beta}\leqno(1.6)$$
Each of these solutions is an admissible ${\bf C}^2$-valued lagrangian distribution in 
the sense of Definition 1.1. Divide all microlocal solutions by the trivial factor $e^{i\pi/4}e^{iE_0\xi_0/h'}$, 
$E_0=E-\alpha x_0$.
Then with the notations of (1.4) the general solution of $({\cal P}^a-E)U=0$ is of the form
$$U=\Sum_{\rho,\e }\alpha_{\e ,\pm\beta}^{(-\nu-1)}
U^{a,-\nu-1}_{\rho,\e ,\pm\beta}\leqno(1.7)$$
The solutions near $a'$ are given by symmetry, e.g. $U^{a',-\nu-1}_{\rho,\e ,-\beta}={}^\vee{\cal I}U^{a,-\nu-1)}_{\rho,\e ,\beta}$.
Moreover both microlocal families can be extended as WKB solutions (satisfying Definition 1.1) along the bicharacteristics.}
\smallskip
Note also that in this region where $\mu(x)$ is a constant,
$U_{\rho,\e ,\pm\beta}=e^{ix\xi_0/h}{\cal U}_{\rho,\e ,\pm\beta;h'}$  with
${\cal U}_{\rho,\e ,\pm\beta;h'}$ oscillating on a frequency scale $1/h'=1/(\alpha h)$,
so if we think of the slope $\alpha$ to be large, 
$U_{\rho,\e ,\pm\beta}$ behaves as a plane wave $e^{ix\xi_0/h}$, modulated by a slow varying function. 
\medskip
\noindent{\bf b) Real holonomy and approximate Bohr-Sommerfeld quantization condition}.
\smallskip
The microlocal kernel $K_h(E)$ of ${\cal P}-E$ on $]-x_0,x_0[\times{\bf R}_+$ can be viewed as a 4-D fibre vector bundle ${\cal F}_h(E)$ 
of admissible Lagrangian distributions over ${\bf S}^1$. We characterize the real part of the resonances
as the set of $E_k(h)$ near $E$ such that this fibre bundle is trivial. We start by computing the holonomy of that bundle.

First we normalize the basis in $K_h(E)$ obtained in Proposition 1.2 
using generalized Wronskians introduced in [HeSj], [Ro].
Namely, let $\chi=\chi^a$ be a smooth cut-off supported on a sufficiently small neighborhood of $a$, equal to 1 near $a$,
$\omega_\pm=\omega_\pm^a$ a small neighborhood of
$\rho_\pm\cap\supp [{\cal P},\chi^a]$, and 
$\chi_{\omega_\pm}=\chi^a_{\omega_\pm}$  a cut-off equal to 1 near $\omega_\pm$. We take Weyl $h$-quantization of these symbols, and
for  $U,V\in K_h(E)$, we call 
$${\cal W}_{\omega_\pm}(U,V)=\bigl(\chi_{\omega_\pm}{i\over h}[{\cal P},\chi]U|V\bigr)=
\bigl(\chi_{\omega_\pm}{i\over h}[{\cal P},\chi]\widehat U|\widehat V\bigr)$$
the {\it microlocal Wronskian} of $(U,\overline V)$ in $\omega_\pm$. 
This is a sesquilinear form on $K_h(E)$, and ${\cal W}_{\omega_\pm}(U,U)$
is independent, modulo error terms ${\cal O}(h^\infty)$, of
the choices of $\chi^a$ and $\chi^a_{\omega_\pm}$ as above. Taking into account both contributions of $\rho_\pm$ we define also
$${\cal W}(U,V)={\cal W}_{\omega_+}(U,V)+{\cal W}_{\omega_-}(U,V)$$
For each microlocal solution $\widehat U=\widehat U^{a,-\nu-1}_{\e ,\pm\beta}$, it turns out that ${\cal W}(U,U)$ 
have asymptotic expansions in $h'$,
of the form $w_0(E,\beta)+h'w_1(E,\beta)+\cdots$, with $w_0(E,\beta)>0$.

Given $\chi=\chi^a$, let now $\widetilde\chi=\widetilde\chi^a$ be a new cut-off equal to 1 on the support of $\chi^a$, and to 0 outside a
slightly larger set.
For $U,V\in K_h(E)$ we set 
$(U|V)_{\widetilde \chi}=(\widetilde\chi U|V)$. Then it is easy to see that there is an orthonormal basis of $K_h(E)$ for the ``scalar product''
$(U|V)_{\widetilde \chi}$, which is at the same time orthogonal for ${\cal W}(U,V)$ (everything being defined modulo ${\cal O}(h^\infty)$~.)
This allows to find $V_{\e }=V^{a,-\nu-1}_{\e ,\beta}$ of the form (1.7) such that $(V_{\e }|V_{\e '})_{\widetilde \chi}=\delta_{\e, \e '}$, 
($\e, \e '=\pm1$),
${\cal W}(V_\pm,V_\pm)>0$, and  ${\cal W}(V_+,V_-)=0$. Of course, by the symmetry
${}^\vee{\cal I}{\cal P}(x,hD_x)={\cal P}(x,hD_x){\cal I}{}^\vee$, such normalized microlocal solutions exist as well near $a'$.
The Lagrangian distributions
$$F^{a,-\nu-1}_{\e ,\beta}=\chi_{\omega^a}{i\over h}[{\cal P},\chi^a]U^{a,-\nu-1}_{\e ,\beta}$$
and similarly $F^{a',-\nu-1}_{\e ,-\beta}$
span the microlocal co-kernel $K_h^*(E)$ of ${\cal P}-E$ in $]-x_0,x_0[\times{\bf R}_+$, as $\e =\pm1$.
The same holds for or $G^{a,-\nu-1}_{\e ,\beta}$ obtained by replacing $U^{a,-\nu-1}_{\epsilon,\beta}$ by the ``orthonormal basis''
$V^{a,-\nu-1}_{\e ,\beta}$ as above.

Because of Proposition 1.2,
the normalized microlocal solutions $V^{a',-\nu-1}_{\e ,-\beta}$ are related to the extension 
of the normalized microlocal solutions $V^{a,-\nu-1}_{\e ,\beta}$ along the bicharacteristics by a monodromy matrix 
$M^{a,a'}=\pmatrix{d_{11}&d_{12}\cr d_{21}&d_{22}\cr}\in U(2)$. Similarly, we obtain $M^{a',a}$ by extending from the left to the right,
and due to symmetry, $M^{a',a}=(M^{a,a'})^{-1}=(M^{a,a'})^*$. Diagonal entries of these matrices are given by action 
integrals along $\rho_\pm$ (see e.g. [Ro]). Off-diagonal terms are ${\cal O}(h')$ and can be computed with the help of the Wronskian
(in the ordinary sense) associated with the system $({\cal P}-E)U=0$ (see e.g. [Ba]). 

The quantization condition is satisfied, precisely when the rank of that system drops of one unit (actually, because of degeneracy,
of 2 units), i.e. when $\dim K_h(E)=\dim K_h^*(E)=2$. This amounts to set to zero the determinant of some Gram matrix $\Gram (E,h)$
expressed in the basis $(V^{a,-\nu-1}_{\e ,\beta}, V^{a',-\nu-1}_{\e ',-\beta})$ and $(G^{a,-\nu-1}_{\e ,\beta}, G^{a',-\nu-1}_{\e ',-\beta})$.
So $E=E_k(h)$ is an eigenvalue, modulo ${\cal O}(h^\infty)$, of ${\cal P}(x,hD_x)$, corresponding to an Andreev state, iff
$\det \Gram (E,h)=0$. 

Here we note the sensitivity of the energy levels $E_k(h)$ with respect to $\phi$. In the ``hard-wall'' limit $\alpha\to\infty$,
we recover the quasi-particle spectrum, of the form $\cos\phi=\cos\bigl({g(E_k(h))\over h}-2\arccos\bigl({E_k(h)\over\Delta_0}\bigr)\bigr)$
for some smooth function $g$ (see [CayMon], [ChLesBl]). 
\medskip
\noindent{\bf 2) The imaginary part of the resonances}.
\smallskip
The considerations above are not sufficient to account for exponentially small corrections to $E_k(h)$.
Further information will be extracted from a Grusin problem.
\medskip
\noindent {\bf a) Microlocal solutions with complex phase}.
\smallskip
Microlocal solutions, computed in the real phase space, are purely oscillating in the metallic lead
$[-x_0,x_0]$. To get information 
in the ``superconducting part of the junction'', we need use 
``infinitesimal'' invariance by time reversal and conjugation of charge.
The substitutions $\xi'\mapsto\pm i\xi'$, or equivalently, $\beta\mapsto\pm i\beta$, leave invariant equation (1.1), with 
a new operator $\widetilde P_{\pm i\beta}$ in Fourier-Laplace representation.
Microlocal solutions of $(\widetilde P_{\pm i\beta}-E)\widetilde u=0$ are constructed similarly and, on the real domain,
independently of those of $(\widetilde P_{\pm \beta}-E)\widetilde u=0$. 

Thus the fibre bundle of microlocal solutions on ${\bf R}\times{\bf R}_+$ (i.e. microlocal kernel of ${\cal P}-E$) splits as
${\cal F}^<_h(E)\oplus{\cal F}_h(E)\oplus{\cal F}^>_h(E)$, 
where we recall ${\cal F}_h(E)$ from Sect.1, and ${\cal F}^{<,>}_h(E)$ are 2-D (trivial) fibre bundles over ${\bf R}$. 

Nevertheless, taking advantage that the coefficients are 
analytic near $a,a'$, there is a way to couple ${\cal F}^{<,>}_h(E)$ with ${\cal F}_h(E)$ in the complex domain. This, together
with the assignment that the global
section be ``outgoing'' at infinity, accounts for complex holonomy.

First we investigate complex holonomy near $a$, and consider the family
of operators, obtained by extending $\widetilde P_{\pm\beta}$ along a path $\{e^{i\gamma}\beta, 0\leq\gamma\leq2\pi\}$
in the complex plane~; similarly, we consider the family
of Lagrangian distributions obtained by extending
$\widetilde u_{\beta}(\xi')$ along that path. They will solve (1.1) iff $\gamma=0,\pm\pi/2,\pi$.

These solutions are related through their Lagrangian manifolds
as follows~: consider (for simplicity) the principal part of $\widetilde P_{\pm \beta}$ and $\widetilde P_{\pm i\beta}$, namely 
$\widetilde Q_{\beta}(-hD_{\xi'},\xi')=(hD_{\xi'})^2+ \bigl(\xi'+\beta\xi'^2\bigr)^2$ and 
$\widetilde Q_{i\beta}(-hD_{\xi'},\xi')=(hD_{\xi'})^2+ \bigl(\xi'+i\beta\xi'^2\bigr)^2$.
The potentials being equal for $\xi'=0$ and $\xi'=-2/(1+i)\beta$, the real Lagrangian manifold
$\rho_+$ near $a$ extends analytically along the loop $\{e^{i\gamma}\beta : \gamma\in[0,2\pi]\}$
in the complex domain, so that it intersects $\rho^>_+$ at $-2/(1+i)\beta$ for $\gamma=\pi/2$.
We can argue similarly for the other branches.
Actually, both $\rho_\pm$ and $\rho^>_\pm$ are branches of a single 2-sheeted Riemann surface, with complex ``turning points''.

We can assign to this analytic manifold microlocal solutions for $\widetilde P_{e^{i\gamma}\beta}$ as in (1.2)
with complex phase, which yields in turn solutions of
$({\cal P}-E)U=0$ for relevant values $0,\pm{\pi\over2},\pi$  of the parameter $\gamma$~; 
these solutions are very similar to the $U^a_{\e ,\pm\beta}$'s given in Proposition 1.2.
The monodromy operator, acting on 
microlocal solutions, is known as {\it connection isomorphism}, see [DeDiPh] and references therein, and also [Fe], 
or [Ro,Sect.4,g] in the case of a system. This connection isomorphism is given by a matrix $N^a\in U(2)$,
whose entries are expressed in term of exponentials of action integrals 
computed along Stokes lines between the complex turning points. 

Let us consider next the conditions at infinity~: for $|x|>x_2$, ${\cal P}$ has constant coefficients, so
we make an analytic dilation of the form $x\mapsto \exp[(\sgn x)\vartheta]\,x$,
$\vartheta>0$. Plane waves with positive momentum have the phase $\exp[ix(\xi_1\pm i\xi_2)/h]$ where 
$\xi_1\pm i\xi_2=\bigl(\mu_0\pm i\sqrt{\Delta_0^2-E^2}\bigr)^{1/2}$, according to the choice of $\rho_\pm^{<,>}$. Analytic distorsion is turned on
for $|x|$ large enough, and $\vartheta$ in the complex upper-half plane. We denote by ${\cal P}_\vartheta$ the distorted operator.
So for all $\im \vartheta\geq0$ small enough, we can make the 
``electronic state'' (resp. ``hole state'') exponentially decaying at $+\infty$ (resp. $-\infty$), which models the 
scattering process $e^+\to e^-$, and similarly for the scattering process $e^-\to e^+$, thus preserving conservation of charge.
\medskip
\noindent {\bf b) A Grusin problem and the width of resonances}.
\smallskip
Following a classical procedure in Fredholm theory, 
we can translate the original eigenvalue problem for ${\cal P}$ into a finite dimensional 
problem {\it via} the Grusin operator [HeSj3,Sect4]~; this is essentially the isomorphism 
$(H^2({\bf R})\otimes{\bf C}^2)/\widetilde K_h(E)\to\Ran ({\cal P}-E)\subset L^2({\bf R})\otimes{\bf C}^2$.
Here $\widetilde K_h(E)$ denotes the 6-D microlocal kernel of ${\cal P}-E$ in ${\bf R}\times{\bf R}_+$, restricted to the set of outgoing 
functions defined above. For ${\cal P}={\cal P}_\vartheta$, we consider ${\cal G}(E)={\cal G}(\vartheta,E)$ of the form~:
$$\eqalign{
{\cal G}(E)&=\pmatrix{&{\cal P}-E&R_-\cr &R_+&0\cr}:
(H^2({\bf R})\otimes{\bf C}^2)\times{\bf C}^6\to (L^2({\bf R})\otimes{\bf C}^2)\times{\bf C}^6\cr
&R_-(x_1,\cdots,x_6)=\Sum_{j=1}^6x_jG_j, \quad R_+U=\bigl((U|G_j)\bigr)_{1\leq j\leq6}\cr
}\leqno(2.1)$$
where the $G_j$'s range over the basis of co-kernel $\widetilde K_h^*(E)$ consisting of $G^a_{\e ,\beta}, G^{a'}_{\e ,-\beta}, G^a_{+,i\beta},
G^{a'}_{-,-i\beta}$ (or their analytic continuation at the branching points). 

At this point we make the following remark~:
Since our Grusin operator (2.1) involves only positive frequencies, it cannot be associated with the self-adjoint operator ${\cal P}_\vartheta$
(for real $\vartheta$).
But resonances are due precisely to a breaking of time-reversal symmetry, and their imaginary part is computed by introducing a
$h$-Pseudo-differential cutoff $\Phi(x,hD_x)$ supported in $\{\xi>0\}$. Because negative frequencies will be eventually removed, we may best think 
of (2.1) as a short-hand notation for the ``full'' Grusin operator ${\cal G}(E)$, that would
take into account the negative frequencies as well. 
 
For all $h>0$ small enough, ${\cal G}(E)$ is bijective, 
with bounded inverse
$${\cal E}(E)=\pmatrix{&{\cal E}_0(E)&{\cal E}_+(E)\cr &{\cal E}_-(E)&{\cal E}_{-+}(E)\cr}$$
and has the property, that $E$ is an eigenvalue of ${\cal P}$ iff $\det {\cal E}_{-+}(E)=0$. 
The construction of ${\cal E}(E)$ is carried as in [HeSj], [Ro], selecting solutions according to the prescriptions above.
Matrix  ${\cal E}_{-+}(E)$ decouples modulo ${\cal O}(h^\infty)$,
with a $4\times4$ block conjugated to $\Gram (E)$~; the interaction with the ``incoming hole'' and ``outgoing electron''
occurs through the ``turning points'' in the complex domain,  
involving the connection isomorphisms $N^a,N^{a'}$. 

For complex $(\vartheta,E)$, we note that $\bigl({\cal P}_\vartheta-E\bigr)^*={\cal P}_{\overline\vartheta}-\overline E$. 
Applying distorsion to the Grusin operator as well, we get~:
$${\cal G}(\vartheta,E)=\pmatrix{&{\cal P}_\vartheta-E&R_-(\vartheta,E)\cr &R_+(\vartheta,E)&0\cr}, \quad 
{\cal E}(\vartheta,E)=\pmatrix{&{\cal E}_0(\vartheta,E)&{\cal E}_+(\vartheta,E)\cr &{\cal E}_-(\vartheta,E)&{\cal E}_{-+}(\vartheta,E)\cr}$$
We can prove that ${\cal G}(\vartheta,E)$ is well-posed for all $\vartheta\in{\bf C}$ small enough, with inverse ${\cal E}(\vartheta,E)$.
Recall from [Ro,Prop.7.1] the following identity~:
\medskip
\noindent {\bf Proposition 2.1}: {\it} Let $\Phi\in C_0^\infty({\bf R}^2;{\bf R})$. With the notations above
$$\eqalign{
\bigl[&R_-(\overline\vartheta,\overline E)^*\Phi{\cal E}_+(\vartheta,E)\bigr]^*{\cal E}_{-+}(\overline\vartheta,\overline E)-
\bigl(\bigr[R_-(\vartheta,E)^*\Phi{\cal E}_+(\overline\vartheta,\overline E)\bigr]^*{\cal E}_{-+}(\vartheta,E)\bigr)^*\cr
&={\cal E}_{+}(\vartheta,E)^*[{\cal P}_{\overline\vartheta},\Phi]{\cal E}_{+}(\overline\vartheta,\overline E)\cr
}\leqno(2.2)$$
In the self-adjoint case, the corresponding statement would be 
``$(R_-^*{\cal E}_+)^*{\cal E}_{-+}$ is self-adjoint''. The determination of the width of resonances then goes as in [Ro],
though it is somewhat more complicated due to the structure of ${\cal E}_{-+}(\vartheta,E)$. 
Take $W(\vartheta,E)\in\ker {\cal E}_{-+}(\vartheta,E)$,
and set ${\cal A}(\vartheta,E)=\bigr[R_-(\vartheta,E)^*\Phi{\cal E}_+(\overline\vartheta,\overline E)\bigr]^*$. From (2.2) and the identity 
$$\bigl(W(\vartheta,E)|{\cal A}(\vartheta,E){\cal E}_{-+}(\vartheta,E)W(\vartheta,E)\bigr)-
\bigl({\cal A}(\vartheta,E){\cal E}_{-+}(\vartheta,E)W(\vartheta,E)|W(\vartheta,E)\bigr)=0$$
we get
$$\eqalign{
\bigl({\cal A}&(\overline\vartheta,\overline E){\cal E}_{-+}(\overline\vartheta,\overline E)W(\vartheta,E)|W(\vartheta,E)\bigr)-
\bigl({\cal A}(\vartheta,E){\cal E}_{-+}(\vartheta,E)W(\vartheta,E)|W(\vartheta,E)\bigr)\cr=
&\bigl({\cal E}_{+}(\vartheta,E)^*[{\cal P}_{\overline\vartheta},\Phi]{\cal E}_{+}(\overline\vartheta,\overline E)W(\vartheta,E)|W(\vartheta,E)\bigr)\cr
}\leqno(2.3)$$
Evaluating both members of this equality gives an implicit equation for the imaginary part of the resonance, showing 
that behaves like
$\exp[-2\int_\tau\xi\, dx/h']$, where $\tau\subset{\bf C}$ is a path connecting the complex branching points in $\rho^>_+\cap\rho_+$.
\medskip
\noindent {\bf References}:
\medskip
\noindent [An] A.Andreev. Zh. Eksp. Teor. Fiz., 46, p.1823 (1964) [Sov. Phys. JETP 19, p.1228 (1964)]

\noindent [Ba] H.Baklouti. Asymptotique des largeurs de r\'esonances pour un mod\`ele d'effet tunnel microlocal.
Ann. Inst. H. Poincar\'e (Physique Th\'eorique) 68(2), p.179-228 (1998)

\noindent [BeIfaRo] A.Bensouissi, A.Ifa, M.Rouleux. Andreev reflection and the
semi-classical Bogoliubov-De Gennes Hamiltonian.
Proceedings ``Days of Diffraction 2009'', Saint-Petersburg. p.37-42. Submitted. 

\noindent [CaMo] J.Cayssol, G.Montambaux. Exchange induced ordinary reflection in a single-channel SFS junction. 
Phys.Rev. B70, 224520 (2004). 

\noindent [ChLeBl] N.Chtchelkatchev, G.Lesovik, G.Blatter. Phys.Rev.B, Vol.62, No.5, p.3559-3564 (2000)

\noindent [DePh] E.Delabaere, F.Pham. {\bf 1}. Exact semiclassical expansions for 1-D quantum oscillators. J.Math. Phys. 38,(12), p.6128-6184 (1997)
{\bf 2}. Resurgence methods in semi-classical asymptotics. Ann. Inst. H.Poincar\'e 71(1), p.1-94, (1999).

\noindent [Fe] M.Fedoriouk. M\'ethodes Asymptotiques pour les Equations Diff\'erentielles Ordinaires. Editions MIR, Moscou (1987)

\noindent [HeSj] B.Helffer, J.Sj\"ostrand. 
{\bf 1}. Analyse semi-classique pour l'equation de Harper. M\'emoire (nouvelle s\'erie) No 3, Soc. Math. de France, 116 (4) (1986).
{\bf 2}. Analyse semi-classique pour l'equation de Harper II.
Comportement semi-classique pres d'un rationnel. M\'emoire (nouvelle s\'erie) No 40, Soc. Math. de France, 118 (1) (1989).
{\bf 3}. Semi-classical analysis for Harper's equation III. M\'emoire  No 39, Soc. Math. de France, 117 (4) (1988)

\noindent [KeSo] J.B.Ketterson, S.N.Song. Superconductivity. Cambridge Univ. Press (1999)

\noindent [ReSi] M.Reed, B.Simon. Methods of Modern Math. Phys. Vol IV. Analysis of Operators. Academic Press (1975)

\noindent [Ro] M.Rouleux. Tunneling effects for $h$-Pseudodifferential Operators,... {\it in}: Evolution Equations, Feshbach
Resonances, Singular Hodge Theory. Adv. Part. Diff. Eq. Wiley-VCH (1999)

\noindent [Sj] J.Sj\"ostrand. Singularit\'es analytiques microlocales, Ast\'erisque No.95 (1982).

\bye